\documentclass[10pt,letterpaper]{article}
\usepackage{aux/macros}
\usepackage[top=0.75in,left=0.65in,right=0.65in,footskip=0.85in]{geometry}

\begin{document}

\begin{center}
    \vspace*{0.5cm}

\Large{Fast, optimal, and dynamic electoral campaign budgeting \\by a generalized Colonel Blotto game} 
 
\vspace*{0.5cm}

\large{Thomas Valles$^{1}$, Daniel Beaglehole$^{2}$}

\vspace*{0.5cm}
\normalsize{$^{1}$Department of Mathematics, UC San Diego.}\\
\normalsize{$^{2}$Computer Science and Engineering, UC San Diego.} 
\vspace*{0.5cm}
\end{center}

\begin{abstract}
    The Colonel Blotto game is a deeply studied theoretical model for competitive allocation environments including elections, advertising, and ecology. However, the original formulation of Colonel Blotto has had few practical implications due to the lack of fast algorithms to compute its optimal strategies and the limited applicability of its winner-take-all reward distribution. We demonstrate that the Colonel Blotto game can be a practical model for competitive allocation environments by implementing the multiplicative weights update algorithm from \textit{Beaglehole et al. (2023)}. In particular, using that this algorithm allows for arbitrary winning-rules, we study strategies for a more realistic model of political campaigning we term \textit{Electoral Colonel Blotto}. Contrary to existing theory and the implemented allocation strategies from U.S. presidential elections, we find that the optimal response to Democratic and Republican strategies in the 2008 and 2020 presidential elections was to focus allocations on a subset of states and sacrifice winning probability on others. We also found that campaigners should compete for undecided voters even in states where the opponent has significantly many more decided voters.
\end{abstract}
\section{Introduction}

The allocation of finite resources, such as advertisements, television appearances, or physical visits, play a pivotal role in the outcomes of elections at all levels \cite{VoterCompositionOutcomes, VoterPersuasionOutcomes}. Further, the volume of resources allocated is increasing with recent elections even after adjusting for inflation. In the United States 2022 midterm elections, both the Democratic and Republican parties spent more than in the past 20 years \cite{MidtermSpending2022}, while in the 2020 presidential and congressional elections, spending totalled \$14.4 billion, doubling the previous record from 2016 \cite{PresidentialSpending2020}. The immensity of these resources demands effective strategies for their distribution.

Despite their volume and influence, it has remained a largely unsolved problem how to optimally allocate these finite resources in electoral campaigns. The Colonel Blotto game has been established as one appropriate theoretical setting to answer this question. In particular, the Colonel Blotto game has been studied extensively as a theoretical model for general competitive allocation scenarios, including political elections \cite{laslier_picard_2002}, advertising and auctions \cite{roberson_2006}, budget allocations \cite{KVASOV2007738}, and even ecological modeling \cite{golman2009}. However, its practical implications have remained limited due to the lack of fast algorithms for computing its optimal strategies. Its practicality has been further limited by the game's all-or-nothing winning rule, which assigns rewards as a step function of your allocation, while practical scenarios call for more general reward functions.

In this work, we enable the Colonel Blotto game to be a practical model for election campaign budgeting, not just a theoretical one. In particular, we introduce a generalization of the Colonel Blotto game, \textit{Electoral Colonel Blotto}, which applies the realistic election model from \cite{three_halves}, and show that one can quickly compute optimal strategies in this game with the multiplicative weights update algorithm of \cite{beaglehole2022sampling}. As a key insight, we show that in the 2008 and 2020 presidential elections, the optimal allocation in response to the historical strategies will frequently sacrifice some states in favor of securing victories in other states. Further, when both players are playing optimally, we observe that proportional allocation is typically close to optimal across winning rules, and not the three-halves strategy, as predicted by the theory in simplified settings and observed empirically in historical election data \cite{three_halves}. We also demonstrate that one should compete for states even if the player needs to sway significantly more than half of the undecided voters. 

More generally, we demonstrate that the MWU algorithm of \cite{beaglehole2022sampling} is state-of-the-art for practical equilibrium computation in Colonel Blotto games. We use this algorithm to resolve Question~1.13 from \cite{beaglehole2022sampling}, concluding that sampled MWU with optimistic updates does not give poly-logarithmic regret in games, as guaranteed by its determinstic variant.

\subsection{Related Work.}

\paragraph{Colonel Blotto} In Colonel Blotto, two or more agents (colonels) allocate resources (soldiers) to shared targets (battles). A colonel wins a battle if they allocate more soldiers than their opponents. The goal of the game is to win as many battles as possible. A number of recent works have developed algorithms to compute exact or approximate solutions to its equilibria \cite{RigolletBlotto, FasterSimplerBlotto, PolynomialBlottoAlgo, ApproximateBlottoFixed}. Prior to these algorithms, many works have constructed exact solutions to the game under restricted settings. See \cite{RigolletBlotto} for an in-depth review of the literature.

\paragraph{Voting strategies} Allocating resources to closely contested states can impact the number of votes won by a candidate \cite{nagler1992presidential} \cite{franz2010political}. A simple resource allocation model which for the American electoral college system was introduced in 1974 \cite{three_halves}. The authors found that the optimal allocation for this model disproportionately favored states which carried more electoral votes. Empirical evidence from the 1960, 1968, and 1976 \cite{bartels1985resource} presidential elections seemed to support this bias to more valuable states, but \cite{colantoni_levesque_ordeshook_1975} argued that the 1960 and 1968 candidates were actually biased toward states which were more closely contested, rather than those which had more electoral votes. \cite{snyder1989election} relaxes some of the assumptions of the model from \cite{three_halves} and found that resources should be allocated to more closely contested states.

A two-stage model in which candidates may allocate resources to states that they are very unlikely to win in the first stage with the hope that they may improve their likelihood of winning in the second stage found that a strategy which targets non-battleground states stochastically is optimal \cite{50state}.

\section{Preliminaries}

\subsection{Standard Colonel Blotto game}
Political campaigns are appropriately modeled by a Colonel Blotto game, where each player is an election candidate seeking to optimally allocate campaign resources. In this model, battles can be, for example, states for presidential elections, where, in each state, securing more votes than the opponent rewards the candidate a certain number of electoral votes. In presidential campaigns, the goal of each candidate is to win more electoral votes in total than their opponent. To model presidential campaigns as a Colonel Blotto game, we consider the related goal of winning as many total electoral votes as possible, in expectation over the randomness of the chosen strategy distributions. The two candidates competitively allocate their campaign resources to achieve this goal.

In the Colonel Blotto Game, two players, players 1 and 2, simultaneously, and without direct knowledge of their opponent's strategy, allocate $n_1$ and $n_2$ troops respectively to $k$ battlefields with values $v_1, \dots, v_k$. The set of possible allocations of each player is the set of $k$-partitions of $n_1$ and $n_2$, which we refer to by $\S_1$ and $\S_2$, respectively. A player wins a battle (and its value) if they allocate strictly more soldiers to that battle than their opponent. In the case of a tied battle, the players split the value of the battle evenly. The total reward earned by a player is the sum of the battle values earned. 

We denote player $1$'s allocation to battle $i$ as $a_i$, and player $2$'s allocation to battle $i$ as $b_i$. In the continuous form of Colonel Blotto \cite{borel1953theory}, a player incurs a loss of $v_i$ if they allocate fewer troops than their opponent to battle $i$ and $0$ otherwise. In the case of ties, i.e., when the players allocate equal soldiers to the same battle $i$, each player incurs a loss of $v_i/2$. In order to guarantee a Nash Equilibrium exists \cite{von1947theory}, players are allowed play distributions over discrete allocations (referred to as \textit{mixed strategies}). The discrete allocations that compose the support of these mixed distributions are called \textit{pure strategies}. Both players seek to minimize their expected loss, where the randomness is over the mixed strategy distributions. Where $\a, \b$ are distributions over the allocations of players 1 and 2, and the discrete allocations $a_1, \ldots, a_k$ and $b_1, \ldots, b_k$ are sampled over these distributions, respectively, the mixed strategy solution to the standard Colonel Blotto is then the solution to the following min-max optimization:
\begin{align*}
    \min_{\a \in \Delta(\S_1)} \max_{\b \in \Delta(\S_2)} \Exp{\a, \b}{\sum_{k' \in [k]}  v_{k'} \cdot \left(\mathbbm{1}_{\{a_{k'} < b_{k'}\}} + 0.5\cdot \mathbbm{1}_{\{a_{k'} = b_{k'}\}}\right)}~,
\end{align*}
where $\Delta(S)$ is the probability simplex supported on the set $S$. Without loss of generality, we consider the zero-sum, centered version of this game in which losses are subtracted by the game's value in the formulation above.

\subsection{Colonel Blotto for electoral campaigning}\label{sec:losses}
The original formulation of Colonel Blotto has limited applicability to political elections, in large part, because elections are not won with probability one by the candidate who devoted more resources to a given battleground. Instead, there is some function (assumed to be known to the players) that determines the probability of winning each battleground of the election based on their and their opponent's allocation to that battleground. 

To generalize Colonel Blotto to political elections, we allow the probability of winning a certain battle to be any bounded function of your allocation, not just a step function as in original Blotto. We refer to the probability of winning a battle given you and your opponent's allocations as the \textit{winning rule}. We define winning rules in terms of probability functions $\ell_{j}(p^{(i)}_j, p^{(\tilde{i})}_j)$, defined as the probability that player $i$ wins battle $j$ if they allocate $p^{(i)}_j$ soldiers to this battle while their opponent, player $\tilde{i}$, allocates $p^{(\tilde{i})}_j$ soldiers to this battle. We now introduce this generalized Blotto game.

\begin{definition}[Electoral Colonel Blotto]
The Electoral Colonel Blotto game is a competition between two players (candidates) labeled $1$ and $2$, each with resource capacities $n_1$ and $n_2$, respectively. These two candidates compete over $k$ shared battles with values $0 < v_1, \ldots v_k < L_{\max}$, where $L_{\max}$ is some positive constant. Where rewards for each battle $j \in [k]$ are allocated according to a winning rule $\ell_{j}: \Delta([n_1]) \times \Delta([n_2]) \goto \Real$. Then, for two mixed strategies $\p^{(1)} \in \mathcal{S}_1$ and $\p^{(2)} \in \mathcal{S}_2$, the reward for players $1$ and $2$, $\mathcal{R}_1(\p^{(1)};\p^{(2)})$ and $\mathcal{R}_2(\p^{(2)};\p^{(1)})$, are given by,
\begin{align*}
    \mathcal{R}_1(\p^{(1)};\p^{(2)}) = \Exp{\p^{(1)}, \p^{(2)}}{\sum_{j \in [k]} v_j \cdot \ell_j (p^{(1)}_j, p^{(2)}_j)} = -\mathcal{R}_2(\p^{(2)};\p^{(1)})~.
\end{align*}
\end{definition}

\subsection{Winning rules for election campaigning}

Various winning rules have been considered to model political elections. One such winning rule is the \textit{Popular Vote} winning rule, which allocates the reward of a battle proportionally to the allocation of a player to that battle. In a more direct model of political campaigns, we also consider the case where players (political candidates) compete over ``undecided voters'' for a battle (e.g. a state in a presidental election). Crucially, while we consider specifically the winning rules studied in prior works, the algorithm of \cite{beaglehole2022sampling} guarantees convergence for arbitrary probability functions ${\{\ell_j : [n_1] \times [n_2] \goto \Real_{\geq 0} \}_{j=1}^k}$, where $[m] = \{1, \ldots, m\}$ for an integer $m \geq 1$. Note the standard Colonel Blotto game is an instance of the Electoral Colonel Blotto, in which the player who devotes more resources to that battle (termed the $0/1$ winning rule) deterministically wins that battle. We formalize these winning rules here for a player $i$, with opponent player $\tilde{i}$,

\begin{itemize}
    \item \textit{0/1 winning rule} (Standard Blotto)- Player $i$ loses the battle $j$ if they allocate fewer soldiers than their opponent. If both players allocate the same amount of soldiers, the winner is determined by a fair coin flip, i.e.,
    \begin{align*}
        \ell_j(p^{(i)}_j, p^{(\tilde{i})}_j) &= v_i \cdot \left(\mathbbm{1}_{\{p^{(i)}_j < p^{(\tilde{i})}_j\}} + 0.5\cdot \mathbbm{1}_{\{p^{(i)}_j = p^{(\tilde{i})}_j\}}\right)
    \end{align*}
    \item \textit{Popular Vote (PV) winning rule} \cite{three_halves}- Player $i$ loses the battle $j$ with probability $\frac{p^{(\tilde{i})}_j}{p^{(i)}_j + p^{(\tilde{i})}_j}$, i.e.,
    \begin{align*}
        \ell_j(p^{(i)}_j, p^{(\tilde{i})}_j) &= v_i \cdot \frac{p^{(\tilde{i})}_j}{p^{(i)}_j + p^{(\tilde{i})}_j}
    \end{align*}
    If both $p_i$ and $\tilde{p}_i$ are $0$, then the loss is $0.5$ for both players. If just $\tilde{p}_i = 0$, then $\ell_j(p_i, \tilde{p}_i) = 1$. 
    
    In the context of presidential elections, the values $v_i$ are understood to be the number of undecided voters in state $i$, and players are trying to maximize the expected number of undecided voters won. In the special case $n_1 = n_2 = n$, the optimal allocation for the popular vote loss is the \textit{proportional allocation}:
    \begin{equation}
    \label{eq:prop_al}
        p_j^* = n \cdot \frac{v_j}{\sum_{j'=1}^k v_{j'}},\quad \forall j \in [k]~.
    \end{equation}
    \item \textit{Electoral Vote (EV) winning rule} \cite{three_halves}- Denote the number of undecided voters in battle (state) $j$ as $u_j = Cv_j$ (where $v_j$ number of electoral college votes for that state), where $C$ is a constant chosen to be even so that $u_j$ is even. An undecided voter votes for player $i$ with probability $\frac{p^{(i)}_j}{p^{(i)}_j + p^{(\tilde{i})}_j}$. Player $i$ loses the state by winning fewer than half of the undecided voters or by winning exactly half of the undecided voters and winning a fair coin flip:
    \begin{align*}
        \ell_j(p^{(i)}_j, p^{(\tilde{i})}_j) &= v_j \cdot \left(\left(\sum_{j' = 0}^{\frac{u_j}{2}-1}\mathbb{P}(X_j = j')\right) + 0.5\cdot \mathbb{P}\left(X_j = \frac{u_j}{2}\right)\right), \quad X_j \sim \text{Binomial}\left(u_i, \frac{p^{(i)}_j}{p^{(i)}_j + p^{(\tilde{i})}_j} \right)~.
    \end{align*}
    If both $p^{(i)}_j$ and $p^{(\tilde{i})}_j$ are $0$, then the state is decided by a fair coin flip - the probability of winning is $0.5$ for both players. 
    
    This setup is slightly different than what is studied in \cite{three_halves}, where a strict majority of undecided voters is required to win a state. Assuming both players have $n$ total resources, a locally optimal allocation for that problem is the \textit{three-halves allocation}:
    \begin{equation}
    \label{eq:th_al}
        p^*_j = n \cdot \frac{v_j^{3/2}}{\sum_j v_j^{3/2}}~.
    \end{equation}
\end{itemize}

For the EV winning rule, we additionally allow for either player to have an advantage in each state. For example, a democratic United States presidential candidate would not need to win half the undecided voters to win a historically democratic state, as there is likely a large group of decided democratic voters. In particular, if player $i$ has an advantage of $\delta_j$ in state $j$, then they lose the state if the win $\frac{u_j}{2}-\frac{\delta_j u_j}{2} - 1$ or fewer undecided voters, and tie if they win $\frac{u_j}{2}-\frac{\delta_j u_j}{2\cdot 100}$ undecided voters. 

\subsection{Learning in Colonel Blotto games}

Machine learning literature offers a relatively simple meta-algorithm for computing solutions to two-player constant-sum games, including Electoral Colonel Blotto. In particular, one can simulate repeated play between the two players, in which strategies are updated in sequential rounds using a learning rule that is guaranteed to give strategies that are not too much worse than the best performing strategy in hindsight \cite{BoostingFreund}. Such an update rule is referred to as a ``no-regret'' learning rule.

\begin{algorithm}
\caption{Repeated play}
\label{alg:meta}
\hspace*{\algorithmicindent} \textbf{Input:} {Players $\{p_\ell\}_\ell$, strategy spaces $\{\X_\ell\}_\ell$, no-regret learning rules $\{f_\ell(\cdot ;t) : \prod_{\ell'} \X_{\ell'}^t \goto \X_\ell\}_\ell$, rounds of play $T$} \\
\begin{algorithmic}[1]
\STATE{$x_\ell^{(0)} \in_R \X_\ell, \quad \forall \ell \in [P]$}
\FOR{$t = 1, \dots, T$} 
\FOR{$\ell = 1, \dots, L$}
\STATE{$x_{\ell}^{(t)} \gets f_\ell(\{x_{\ell'}^{(t')}\}_{t'\in [t-1], \ell' \in [L]})$}
\ENDFOR
\ENDFOR
\end{algorithmic}
\end{algorithm}

The repeated play dynamic is a randomized optimization procedure for solving zero-sum games. The performance of a learning rule in repeated play optimization is measured by its \textit{regret}. The regret of player $i$'s average allocation in iteration $T$ is defined as,
\begin{equation*}
    \text{Regret}_i(T) = \frac{1}{T}\,\sum_{t = 1}^T\sum_{j = 1}^k \ell_{j}^{t}\left(p^{(i,t)}_j, p^{(\tilde{i},t)}_j\right) - \min_{\alpha \in S_i} \frac{1}{T} \sum_{j = 1}^k L_i^{(T)}\left(j, \alpha_{j}\right)
\end{equation*}
where $p^{(i,t)}_j$ is player $i$'s allocation to battle $j$ in iteration $t$. That is, the regret of a player is how much more loss the player incurred compared to the single best strategy in hindsight. 

We bound the distance to convergence of this optimization procedure by the sum of the \textit{regret} values of the two-players (Proposition~\ref{prop:regret_opt}). 

\begin{proposition}\label{prop:regret_opt}
The approximation error of Algorithm~\ref{alg:meta} is bounded by the sum of the regrets of the two players.
\end{proposition}

We refer to this sum as the \textit{total regret} of this pair of strategies. Note that the distance to equilibrium, the exact quantity being optimized, can be computed directly. However, this quantity naïvely requires computation that is quadratic in the number of rounds played, in contrast to the total regret bound which can be computed with a linear dependence on $T$. Note, however, the total regret in the cases we study is quite loose compared to the true distance to equilibrium (Figure~\ref{fig:reg_vs_eq}). 

The no-regret learning rule we apply here is Multiplicative Weights Update (MWU) \cite{MultiplicativeWeightsUpdate}. In this algorithm, the player maintains a value (called a weight), for each of their potential strategies, and updates individual weights based on the total performance of the corresponding strategy over all rounds of repeated play. In particular, the weights are initialized uniformly, then updated according to the exponentiated performance (loss). We describe MWU in Algorithm~\ref{alg:mwu}. Note the loss functions and weights after the initialization are random variables, as strategies are sampled from the distributions given by MWU.  

\begin{algorithm}
\caption{Multiplicative Weights Update (MWU)}
\label{alg:mwu}
\hspace*{\algorithmicindent} \textbf{Input:} {Set of strategies $\S$, loss functions $\{\ell_{t} : \S \rightarrow \Real_{\geq 0}\}_{t=1}^T$, learning rate $\beta \in (0,1)$} \\
\begin{algorithmic}[1]
\STATE{Initialize a weight per strategy: $w_0(\x) \gets \frac{1}{|\S|},\quad \forall\, \x \in\S$}
\FOR{$t = 1, \dots, T$}
\STATE{$w_{t}(\x) \gets w_{t-1}(\x) \cdot \beta^{\ell_{t}(\x)},\quad \forall\, \x \in\S $}
\ENDFOR
\end{algorithmic}
\end{algorithm}

While the MWU algorithm is in general implemented explicitly, where a single value is indeed stored for every strategy, the algorithm of \cite{beaglehole2022sampling} uses that these loss functions that appear in Colonel Blotto games are highly structured, and one may implicitly access such weights as if they were stored explicitly when sampling strategies. This technique allows the user to implement MWU, even when the number of strategies, $|S|$, is too large to write down. Furthermore, in contrast to other implementable algorithms for equilibrium computation in Colonel Blotto in general settings, this algorithm provides the following fast convergence guarantee in the Electoral Blotto games.

\begin{proposition}[Fast MWU in Blotto games\cite{beaglehole2022sampling}]
    We can sample MWU from Electoral Colonel Blotto with $n$ soldiers, $k$ battles with any winning rule with bounded (constant) rewards in $O(nk)$ time per round of play. This gives complexity to compute an $\varepsilon$-approximate Nash equilibrium, for $\varepsilon > 0$, in time $O(nk^2 \varepsilon^{-2})$ with high (constant) probability. 
\end{proposition}

In our optimizations with repeated play, the accumulated loss over time for player $i$ is stored in the historical loss matrix $L_i^{(t)} \in \R^{k\times (n_i+1)}$, where $L_i^{(t)}(j, m)$ denotes the loss of allocating $m$ soldiers to battle $j$ accumulated through $t$ iterations of the algorithm. We allow for the matrix $L_i^{(0)}$ to be initialized with a ``warm start'', having the loss for each allocation if the opponent had played a uniform allocation (each battle gets the same number of soldiers), proportional (equation~\ref{eq:prop_al}), or three-halves allocation (equation~\ref{eq:th_al}) for $I_0$ rounds. This initialization technique allows us to study whether the dynamics of repeated play diverge from the suggested equilibria in prior works.
\section{Results}

We present our main insights for strategies of the Electoral Colonel Blotto game and fast equilibrium computation with MWU.

\subsection{Presidential campaigning as Colonel Blotto}

\paragraph{Proportional allocations are often close to optimal across winning rules.}

\begin{figure}[!htbp]
    \centering
    \includegraphics[width = 14cm]{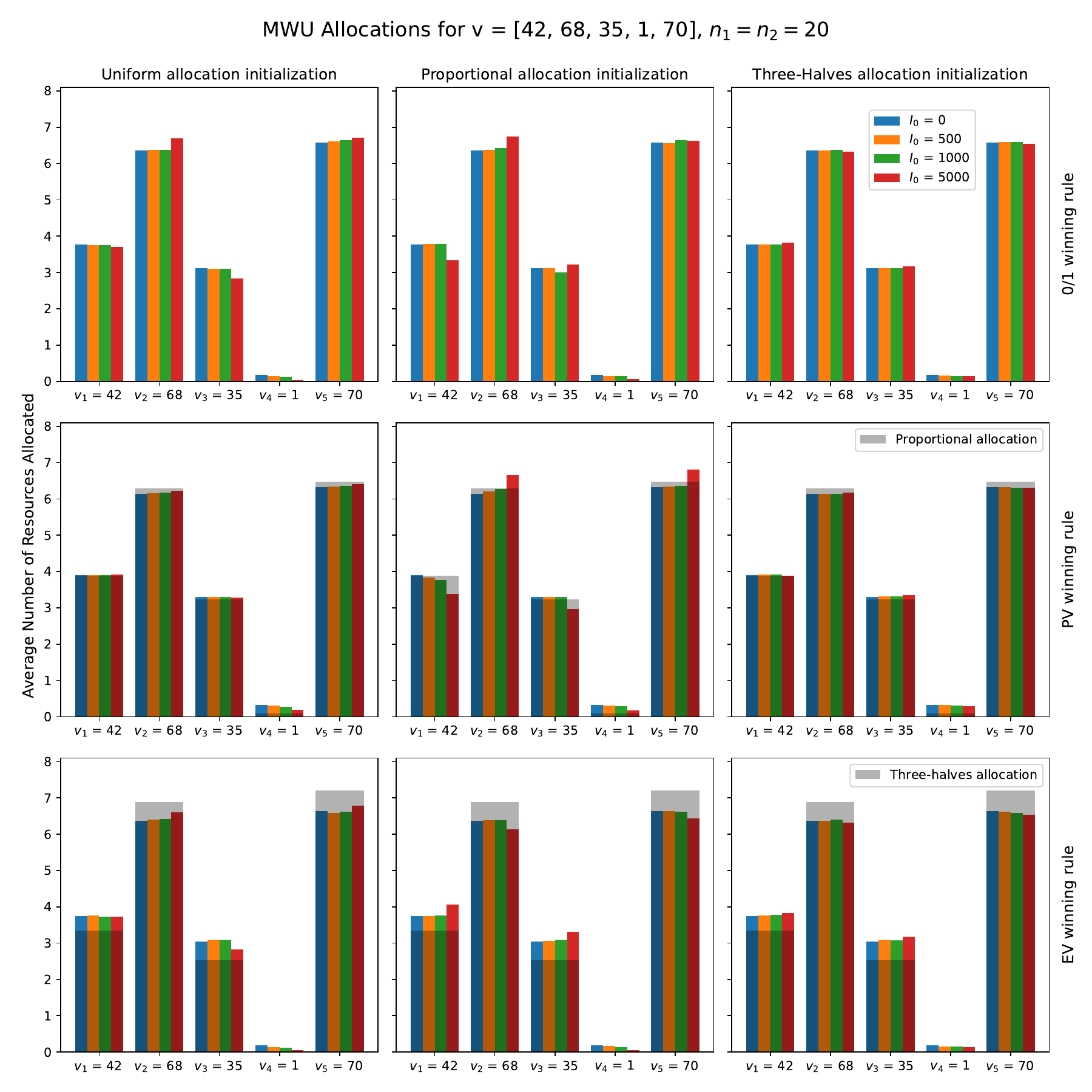}
    \caption{Allocations from MWU for battles with symmetric values, and equal resource capacities for the two players. Columns indicate the choice of warm-start strategy for the historical loss matrix, and $I_0$ indicates the number of warm-start rounds. Bar heights represent the average number of soldiers allocated by the one such player under the $0/1$, Popular Vote (PV), and Electoral Vote (EV) winning rules. For the Popular Vote (PV) and Electoral Vote (EV) winning rules, the gray bars indicate the the theoretical equilibria proposed in \cite{three_halves}. Battle values were chosen uniformly at random from $\{1,\ldots,100\}$.}
    \label{fig:allocations_not_optimistic_rand}
\end{figure}

\begin{figure}[!htbp]
    \centering
    \includegraphics[width = 14cm]{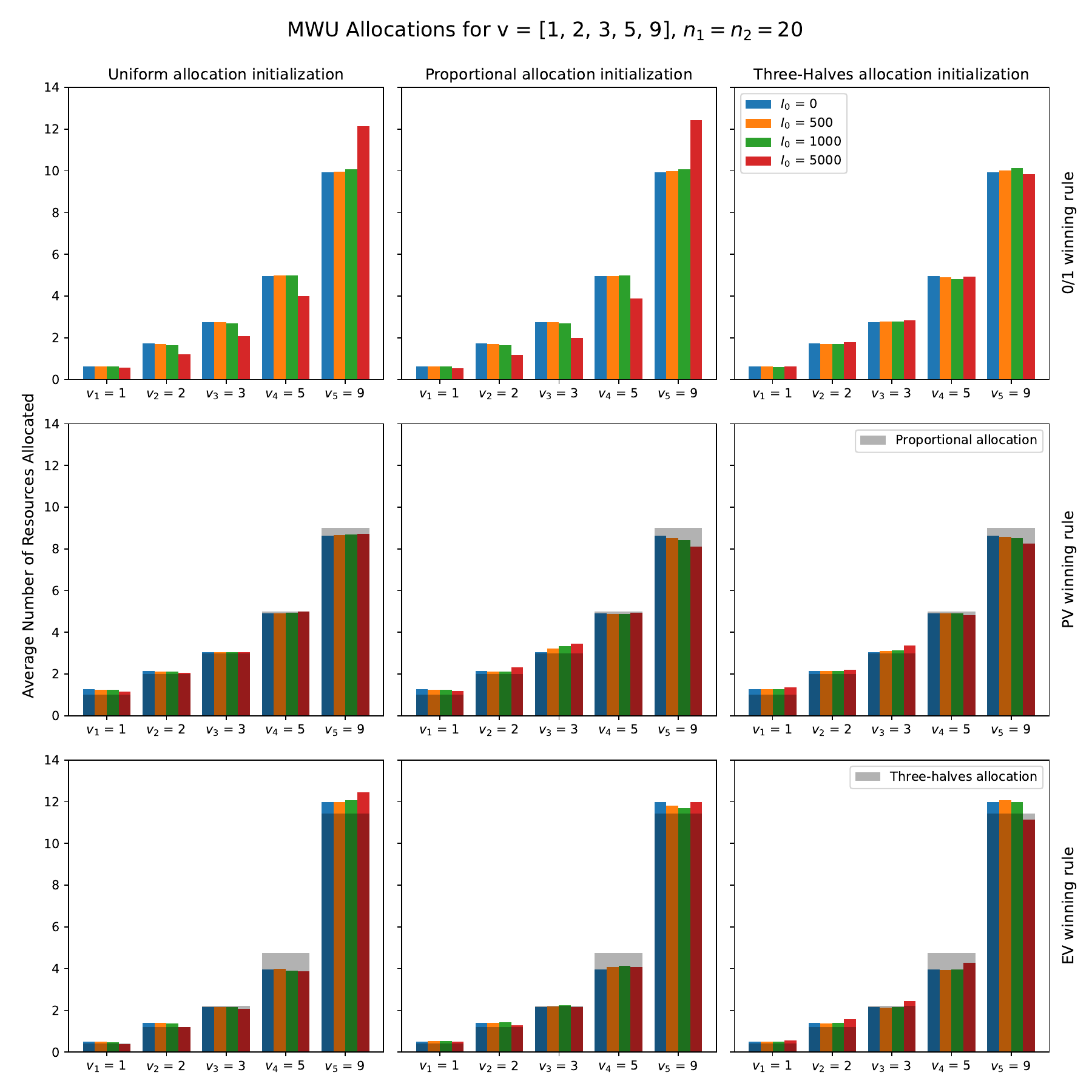}
    \caption{Allocations from MWU for battles with symmetric values, and equal resource capacities for the two players. Columns indicate the choice of warm-start strategy for the historical loss matrix, and $I_0$ indicates the number of warm-start rounds. Bar heights represent the average number of soldiers allocated by the one such player. For the Popular Vote (PV) and Electoral Vote (EV) winning rules, the gray bars indicate the the theoretical equilibria proposed in \cite{three_halves}. }
    \label{fig:allocations_not_optimistic_fixed}
\end{figure}

We first verify that the MWU allocations for the PV winning rule match the proportional allocation predicted in \cite{three_halves}, and described in \ref{sec:losses}, (middle rows of Figures~\ref{fig:allocations_not_optimistic_rand} and \ref{fig:allocations_not_optimistic_fixed}), across battle value distribution and strategy initialization schemes. As expected of the PV vote, in both sets of battle values (Figures \ref{fig:allocations_not_optimistic_rand} and \ref{fig:allocations_not_optimistic_fixed}) the MWU allocation for the 0/1 winning rule is very close to the proportional allocation, across initilization schemes. Interestingly, optimal strategies for the 0/1 winning rule are also roughly equal to the proportional allocation. 

In contrast to the prediction of \cite{three_halves}, the MWU allocation for the electoral vote winning rule is much closer to the proportional allocation than the three-halves allocation when battle values are sampled uniformly. To measure distances between mixed strategies we represent these distributions by their marginals over battles. I.e., for a distribution $\pi \in \Delta(\mathcal{S})$ over the set of potential strategies $\mathcal{S}$, we represent $\pi$ by the vector of its marginal distributions $(\Exp{\x\sim\pi}{x_1}, \ldots, \Exp{\x\sim\pi}{x_k})$, where $\x$ is a random allocation allocation sampled from $\pi$. With the uniformly sampled battle values, the electoral vote winning rule, and without using a warm start for strategies ($I_0 = 0$, Figure \ref{fig:allocations_not_optimistic_rand}), we found that the Euclidean distance between the proportional allocation and the optimized mixed strategy is $0.501$ versus a distance of $0.828$ for the three-halves allocation (the total regret for the optimized mixed strategy is $0.023$). Note this difference remains even when initializing the game history to $5000$ rounds of exactly the three-halves allocation, suggesting that the game dynamics diverge from this strategy choice. However, when battle values are increasing quadratically from one battle to the next (Figure \ref{fig:allocations_not_optimistic_fixed}), the optimized mixed strategy is indeed closer to the three-halves allocation than the proportional allocation (the distances of the optimal solution from the three-halves and proportional strategies are $3.350$ and $0.963$, respectively). This suggests that the imbalanced regime is where the approximation of \cite{three_halves} becomes more accurate.

\begin{figure}[!htbp]
    \centering
    \includegraphics[width = 14cm]{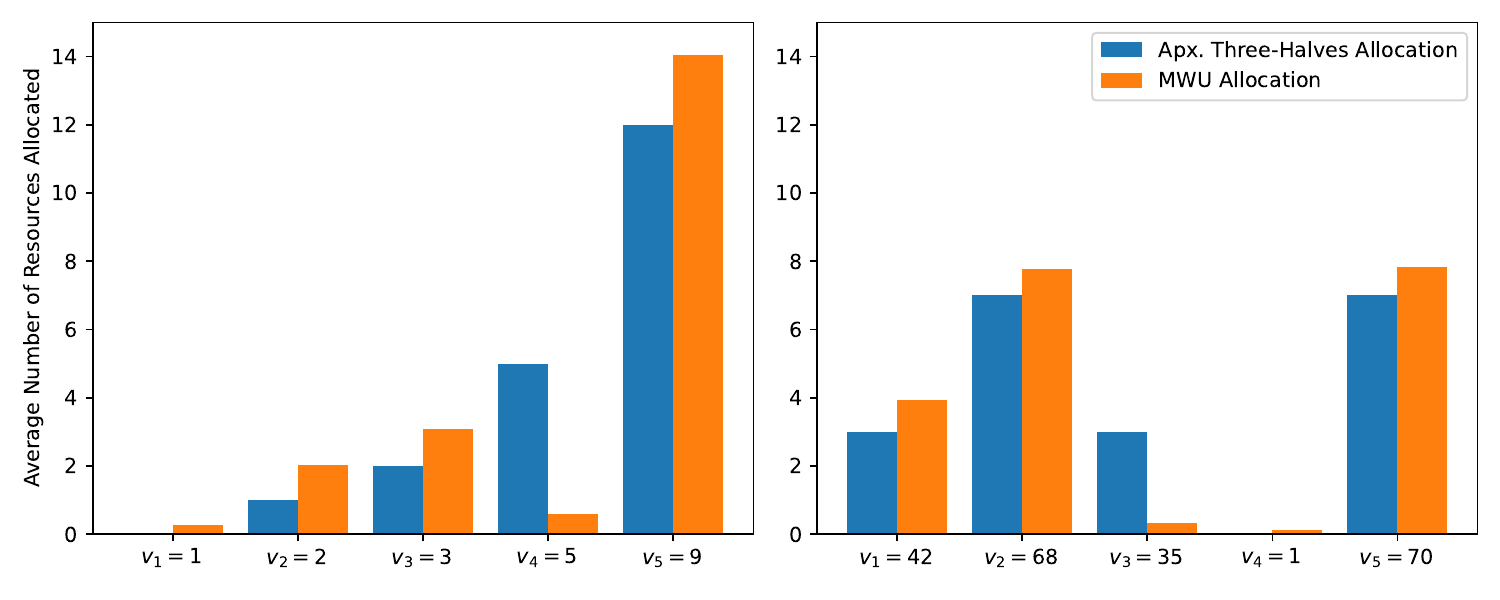}
    \caption{Comparison of regret and average allocation when one player plays an integer approximation of the Three-Halves Allocation while the opponent plays the MWU allocation. We use the Electoral Vote winning rule with $I_0 = 0$ for the MWU allocation.}
    \label{fig:deterministic_allocs}
\end{figure}

\paragraph{Exploitability of the three-halves allocation.}
It has been empirically observed in presidential campaigns that candidates will often allocate resources according to the three-halves allocation \cite{three_halves}. This allocation is indeed an equilibrium in the Electoral Vote winning rule, provided the two candidates \cite{three_halves} distribute equal allocations to all states. However, we demonstrate that the three-halves allocation is heavily exploitable, when this restricted condition does not hold, by \textit{sacrificing} battles in both synthetic and historical presidential elections. In the synthetic setup, we apply the optimal Algorithm~\ref{alg:BlottoMWU} against the three-halves allocation, and verify the convergence of the MWU algorithm in Figure~\ref{fig:determinist_th_regrets}. In both scenarios, the MWU player sacrifices victory, allocating few resources, on one battle of moderate value, and focuses their resources on winning one or more valuable battles (Figure~\ref{fig:deterministic_allocs}). 

\begin{figure}[!htbp]
    \centering
    \includegraphics[width = 17cm]{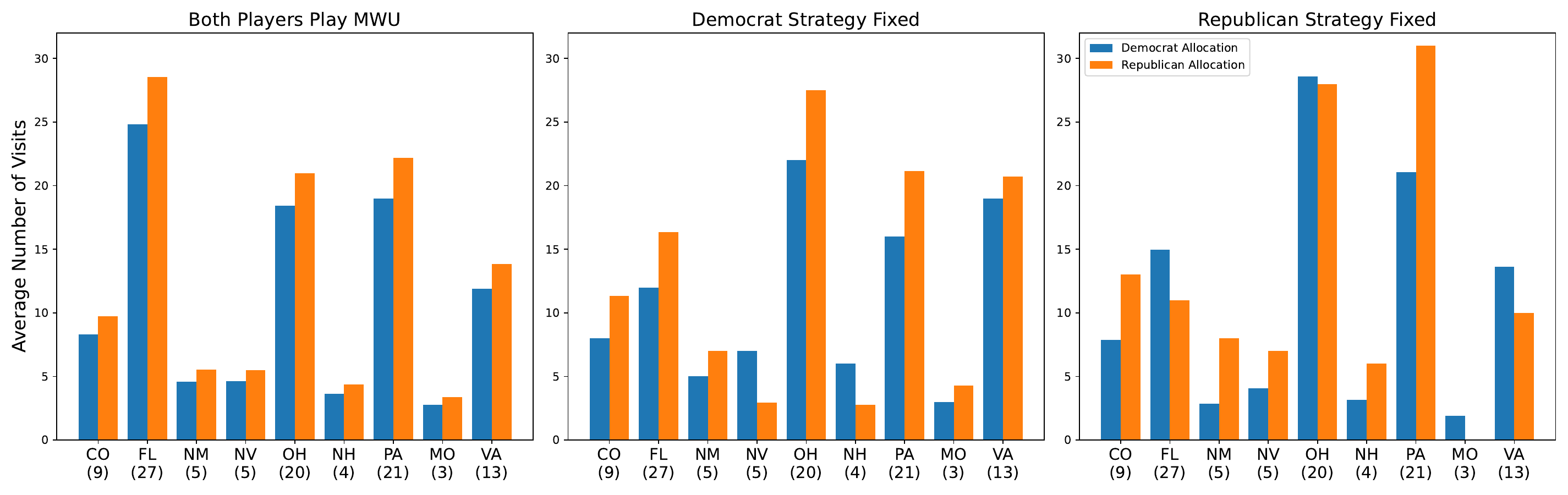}
    \caption{Average allocations returned from the MWU algorithm  using data from the 2008 presidential candidates. States presented are those which were listed as being viewed as battleground states by at least one of the two parties according to \cite{2008_strategies}. The fixed strategies represent the actual number of visits made by the ticket \cite{2008_strategies}. Electoral college votes for each state at the time of the election are in parenthesis.} 
    \label{fig:2008_election_allocs}
\end{figure}

We observe a similar phenomenon on historical election data. In 2008, there were nine states which were considered battleground states by at least one of the two parties \cite{2008_strategies}. The Republican ticket had $114$ visits in total to these battleground states, while the Democratic ticket had only $98$. We run the MWU algorithm without a warm start and with the Electoral Vote winning rule. We consider three learning rule choices for the two players:
\begin{itemize}
    \item Both parties use the optimal MWU algorithm.
    \item The Democratic ticket plays a fixed strategy (i.e. they always use the allocation corresponding to the number of visits to the battleground states presented in \cite{2008_strategies}) while the Republican ticket uses the MWU algorithm.
    \item The Republican ticket plays a fixed strategy while the Democratic ticket uses the MWU algorithm.
\end{itemize}  
As implied by the no-regret bound of the MWU algorithm, when one party is allowed to adapt to the fixed strategy of their opponent, the regret of this player decreases through repeated play. Interestingly, the fixed historical strategies played during the presidential campaigns perform poorly against the optimized strategies, giving half the maximum loss (Figure~\ref{fig:2008_election_regrets}). Consistent with this observation, we see that the marginal distributions over strategies returned by MWU against a fixed or optimal opponent, for either party, are far from the historical allocations (Figure~\ref{fig:2008_election_allocs}). In adapting their visit allocations, we observe that both parties tended to sacrifice less valuable states to guarantee victory in more valuable ones. For example, when the Republican strategy is fixed, the Democratic ticket allocated more resources on average to four of the nine battleground states, including three of the four most valuable states in terms of electoral college votes, while allocating very few visits to the remaining ones (Figure \ref{fig:2008_election_allocs}). 

\begin{figure}[!htbp]
    \centering
    \includegraphics[width = 17cm]{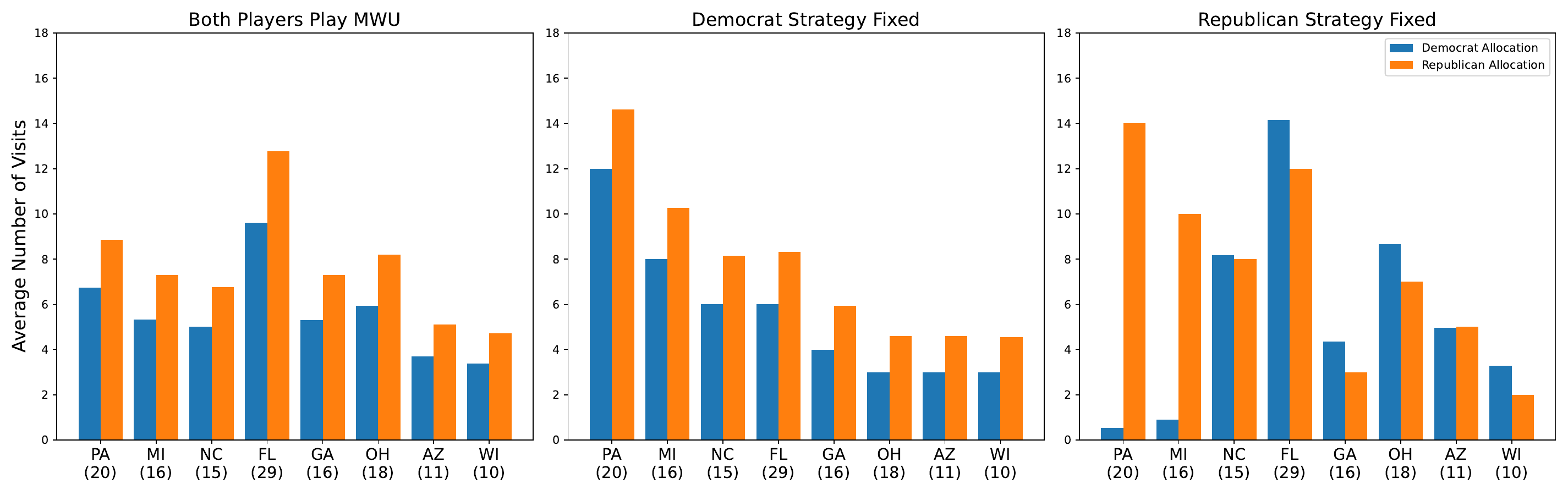}
    \caption{Resource allocations using data from the 2020 presidential candidates. Biden and Harris had $45$ total visits, while Trump and Pence had $61$. States presented are those which were visited at least twice by each ticket \cite{2020_strategies}. Electoral college votes for each state at the time of the election are in parenthesis.}
    \label{fig:2020_election_allocs}
\end{figure}

For the 2020 election, sacrificing was again the optimal strategy against a fixed opponent. In this case, the Democratic ticket, who again had fewer total visits than the Republicans, allocated more than the Republican ticket on average in five of the eight battleground states, while almost entirely sacrificing the remaining states (Figure~\ref{fig:2020_election_allocs}). In contrast, the Republican ticket did not need to sacrifice, as they had sufficiently many total visits to guarantee victory in all states. Note for this election, to reduce the total number of battles, we considered the eight states where each ticket visited at least twice.  

\begin{figure}[!htbp]
    \centering
    \includegraphics[width = 14cm]{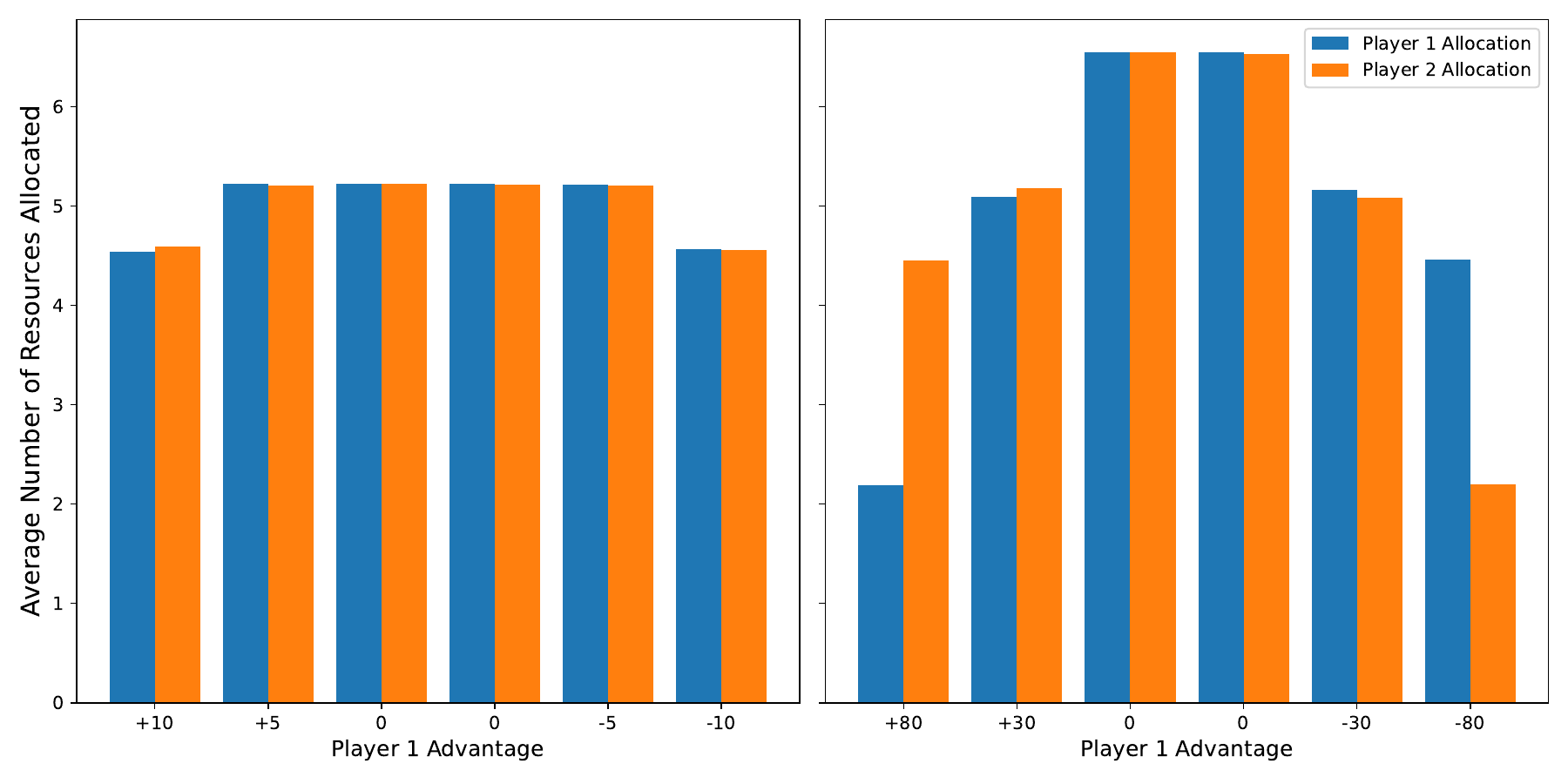}
    \caption{Averaged allocations for two players. Each battle is equally valued. player $1$'s advantage in each battle is listed below. An advantage of $+80$ means that player $1$ requires $80\%$ fewer undecided voters than player $2$ to win the state.  $n_1 = n_2 = 30$.}
    \label{fig:adv_test}
\end{figure}

\paragraph{Competing for states where the opponent has an advantage.} Empirical analysis of campaign spending in the 1960 and 1968 U.S. presidential elections suggested that candidates allocated more resources to states that are more closely contested \cite{colantoni_levesque_ordeshook_1975}. We tested to see if this is an optimal strategy by running our algorithm with six uniformly weighted battles and two sets of advantages (Figure \ref{fig:adv_test}). The average allocation favored battles that are closer, but the difference was only apparent when the advantages were large. The disadvantaged player allocated more resources than the advantaged player in battles that are not evenly contested.

\subsection{State-of-the-art equilibrium computation for Colonel Blotto} \label{sec:effiency}

We demonstrate that the sampled MWU Algorithm~\ref{alg:BlottoMWU} is the, empirically, state-of-the-art algorithm for computing equilibria in Colonel Blotto games. Full experimental results are provided in Section~\ref{sec:algo_speed}. We observe that this algorithm terminates in less time than the only other (implemented) practical algorithm for Colonel Blotto games \cite{FasterSimplerBlotto}. The difference becomes more significant as the number of soldiers for each player increases. For instance, with $k = 20$ battles and $n_1 = n_2 = 30$ soldiers, Algorithm~\ref{alg:BlottoMWU} requires around $17.6$ seconds to achieve a $0.05$-approximate Nash equilibrium, while the algorithm used in \cite{FasterSimplerBlotto} requires $9$ minutes and $13$ seconds for exact computation. Full runtime results across battle and solider settings are given in Section~\ref{sec:algo_speed}.

Further, we use our fast MWU implementation to resolve Question 1.13 in \cite{beaglehole2022sampling} -  sampling MWU with \textit{optimistic} updates does not give poly-logarithmic regret (as a function of time $T$) for Colonel Blotto games, even though its deterministic variant does \cite{optimstic_log_regret}. In particular, one can straight-forwardly implement Algorithm~\ref{alg:BlottoMWU} with optimistic updates that double weight the loss of the most recent round and subtract off the loss of the second-most recent round. This reflects the optimistic intuition that more recent losses are likely to be more informative than older ones \cite{optimistic_hedge}. Note, however, that the deterministic implementations of MWU and its optimistic variant lose the computational efficiency of sampling, and appear intractable. See Section~\ref{sec:standard vs. optimistic} for the full experimental results.

\section{Discussion}

Finding an optimal allocation of finite resources such as television ad time or campaign visits can be viewed through the lens of the well-studied Colonel Blotto game. Despite the importance of the game's applications and the myriad literature on the topic, efficient and practical algorithms for equilibrium computation in Colonel Blotto have not been developed. 

We demonstrate that the MWU algorithm of \cite{beaglehole2022sampling} offers an effective and practical solution to solving Colonel Blotto games with general winning rules and apply the algorithm to derive specific insights into electoral campaign strategies. The MWU algorithm is guaranteed to be no-regret against any adversary and to converge to a Nash Equilibrium as long as the winning rule is such that the game is zero-sum. Our implementation of the algorithm allows us to find approximate equilibria significantly faster than other previous studied approaches, in more general settings, and at scales previously intractable for other algorithms. Further, modifications to the winning rules discussed in Section \ref{sec:losses} can be easy to implement, and more sophisticated winning rules that model election procedures other than the US presidential election may also be possible to put into practice. 

\subsection{Next steps}

\paragraph{Political and economic modeling.} A number of interesting and important questions in modeling campaign allocations arise from this work. For example, one could develop a more comprehensive election model that better accounts for (1) natural advantages of a player in some states, (2) the varying degrees to which an undecided voter may be swayed by additional resources, (3) the cost of deploying additional resources to different states, or (4) the possibility of investing resources in a state that a candidate is unlikely to win in the hopes of flipping the state in a future election. This work also demonstrates that the MWU algorithm may be an effective method for the diverse set of applications modeled by the Colonel Blotto game, including allocating funds in stock portfolios \cite{MWU_stocks} and learning to bid in advertising auctions \cite{AuctioGym}.

\paragraph{Distilling mixed strategies into pure ones.} While the approximate equilibria solutions returned by MWU are non-discrete, many practical resource allocation scenarios permit playing just a single allocation. This can be problematic if one is using the algorithm to find the optimal allocation of a resource that is discrete in nature (e.g. choosing a single set of campaign visits). If a discrete allocation is required, one possibility would be to deterministically play the single pure strategy that was most frequently played by the algorithm (the mode of your optimized distribution). Instead of just a single strategy, we may further allow players to choose a small set of strategies. Choosing the best subset to represent the entire optimized distribution given by MWU is a computing problem known as distribution compression, and has been studied in prior work \cite{dist_compression}. It would be useful to see if the structure of the MWU distribution lends to more specific algorithmic compression approaches.

\paragraph{Limited bit precision when $T$ is large.} At the technical level, we are unable to run the optimization algorithm to arbitrarily small regret due to the bit complexity of the stored weights. In particular, after $T$ rounds of MWU with learning rate $\beta$ and maximum total loss $\Lmax$, the smallest number that would need to be stored on memory is $\beta^{T\Lmax}$. Increasing the learning rate $\beta$ closer to $1$ can increase this minimum value, though more iterations will be required to achieve a desired optimization error.  Developing a technique to allow for longer optimization periods with non-vanishing learning rates would lead to better equilibria approximation. 

\newpage
\bibliography{aux/refs}{}
\bibliographystyle{abbrv}

\newpage
\appendix
\section{Implementation details}

\begin{figure}[!htbp]
    \centering
    \includegraphics[width = 14cm]{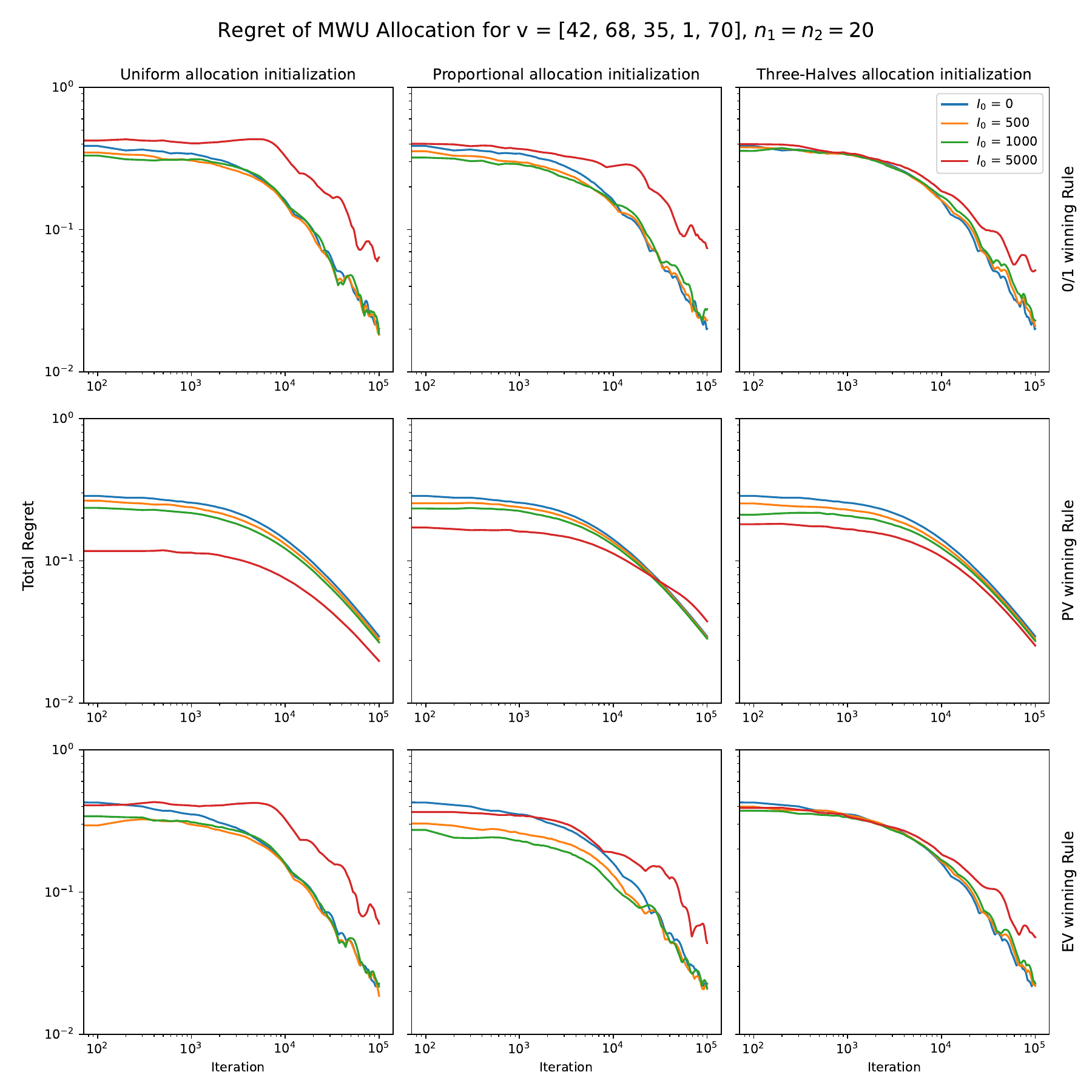}
    \caption{Total regret for the MWU allocations from Figure \ref{fig:allocations_not_optimistic_rand} using a log-log scale. The horizontal axis in each panel represents the round of repeated play as detailed in Algorithm \ref{alg:meta}.}
    \label{fig:regrets_not_optimistic_rand}
\end{figure}

\begin{figure}[!htbp]
    \centering
    \includegraphics[width = 14cm]{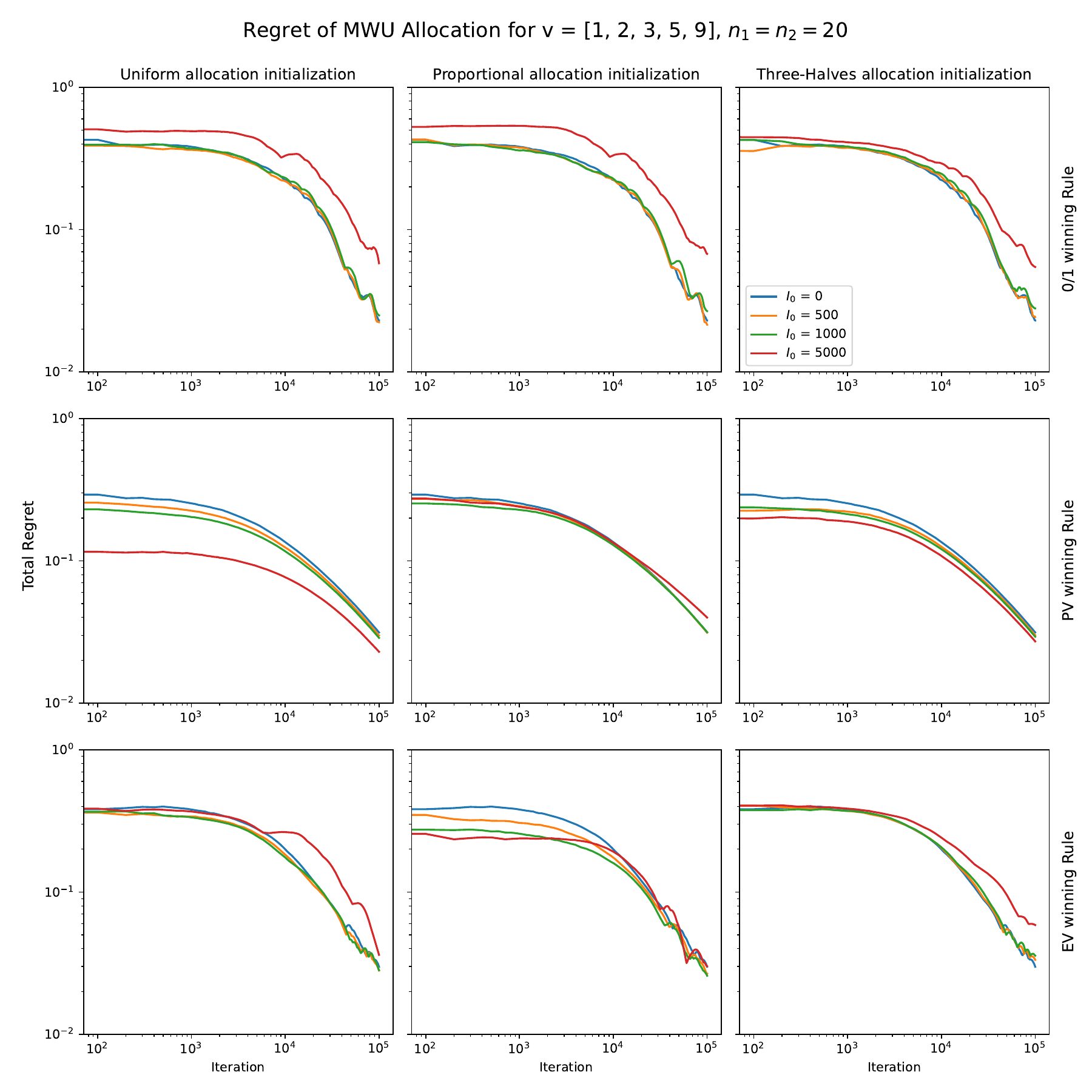}
    \caption{Total regret for the MWU allocations from Figure \ref{fig:allocations_not_optimistic_fixed} using a log-log scale. The horizontal axis in each panel represents the round of repeated play as detailed in Algorithm \ref{alg:meta}.}
    \label{fig:regrets_not_optimistic_fixed}
\end{figure}

\begin{figure}[!htbp]
    \centering
    \includegraphics[width = 14cm]{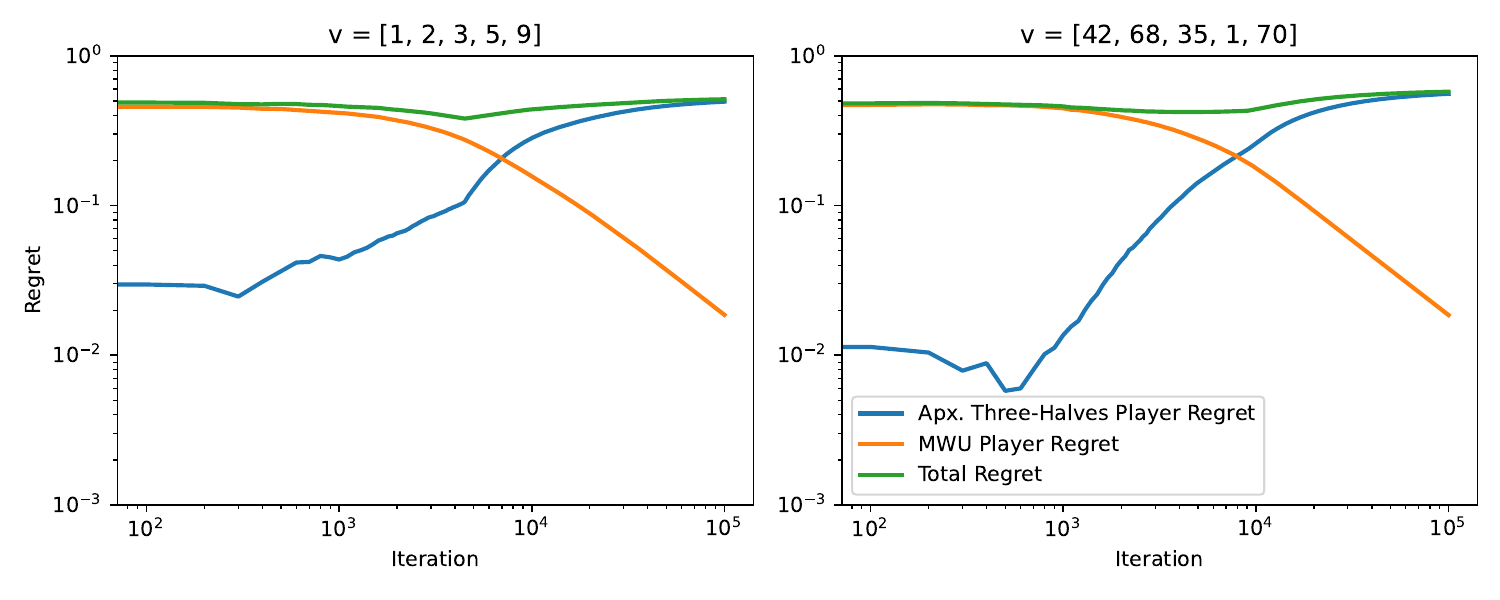}
    \caption{Total regret for the MWU allocations from Figure \ref{fig:deterministic_allocs} using a log-log scale. The horizontal axis in each panel represents the round of repeated play as detailed in Algorithm \ref{alg:meta}.}
    \label{fig:determinist_th_regrets}
\end{figure}

\begin{figure}[!htbp]
    \centering
    \includegraphics[width = 14cm]{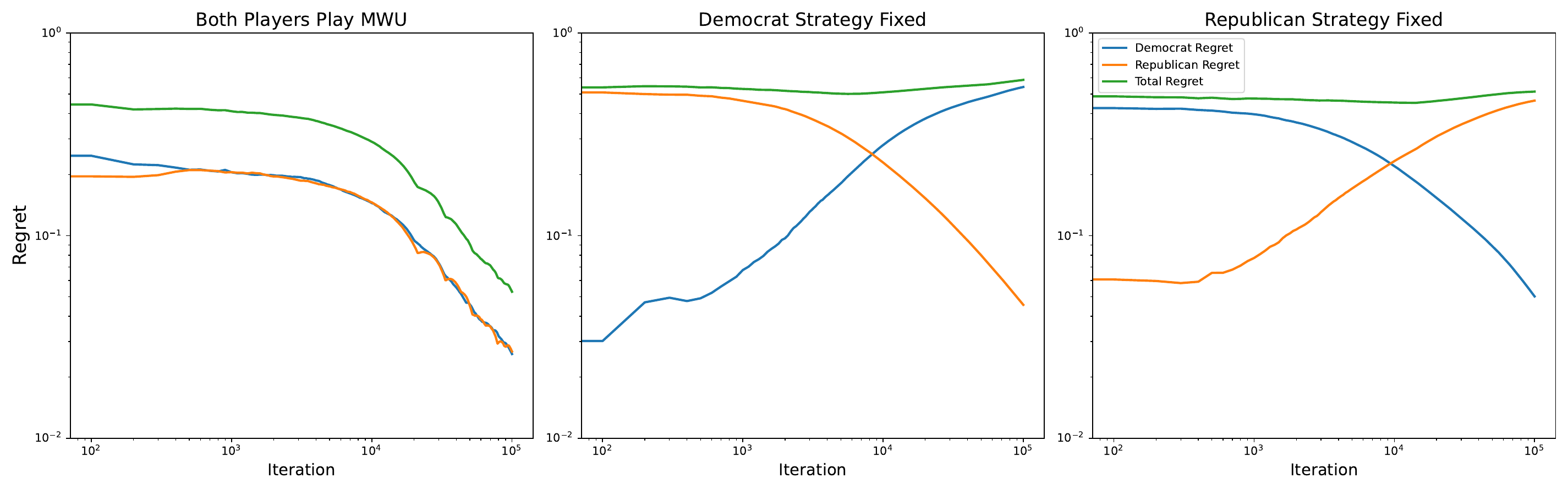}
    \caption{Total regret for the MWU allocations from Figure \ref{fig:2008_election_allocs} using a log-log scale. The horizontal axis in each panel represents the round of repeated play as detailed in Algorithm \ref{alg:meta}.}
    \label{fig:2008_election_regrets}
\end{figure}

\begin{figure}[!htbp]
    \centering
    \includegraphics[width = 14cm]{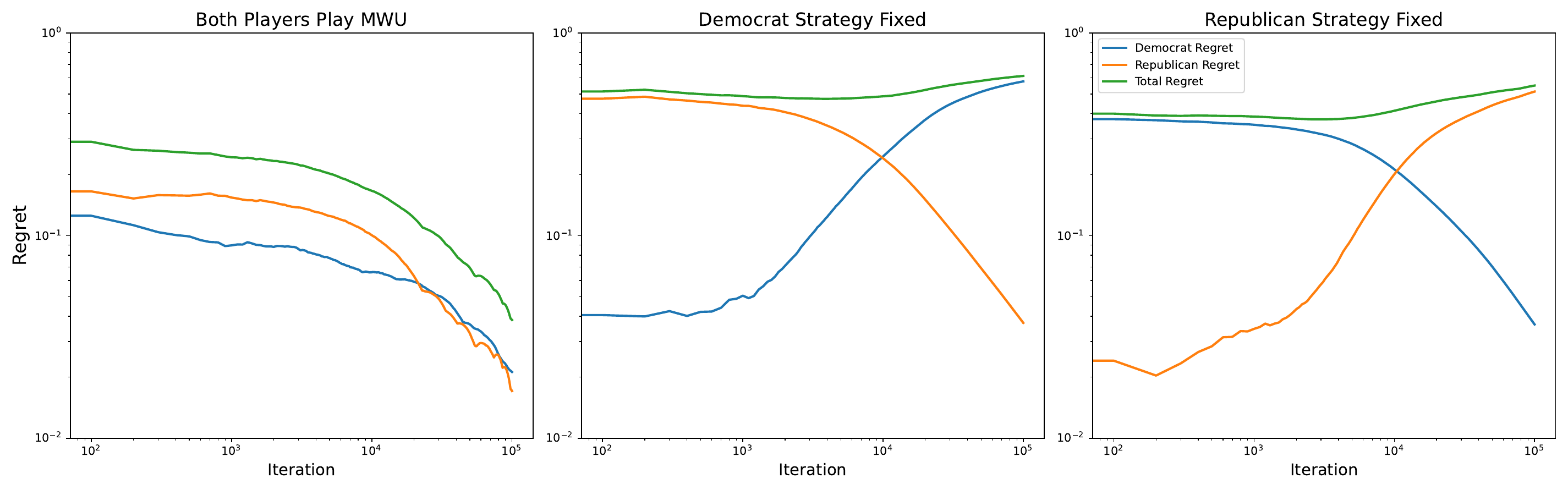}
    \caption{Total regret for the MWU allocations from Figure \ref{fig:2020_election_allocs} using a log-log scale. The horizontal axis in each panel represents the round of repeated play as detailed in Algorithm \ref{alg:meta}.}
    \label{fig:2020_election_regrets}
\end{figure}

\begin{figure}[!htbp]
    \centering
    \includegraphics[width = 14cm]{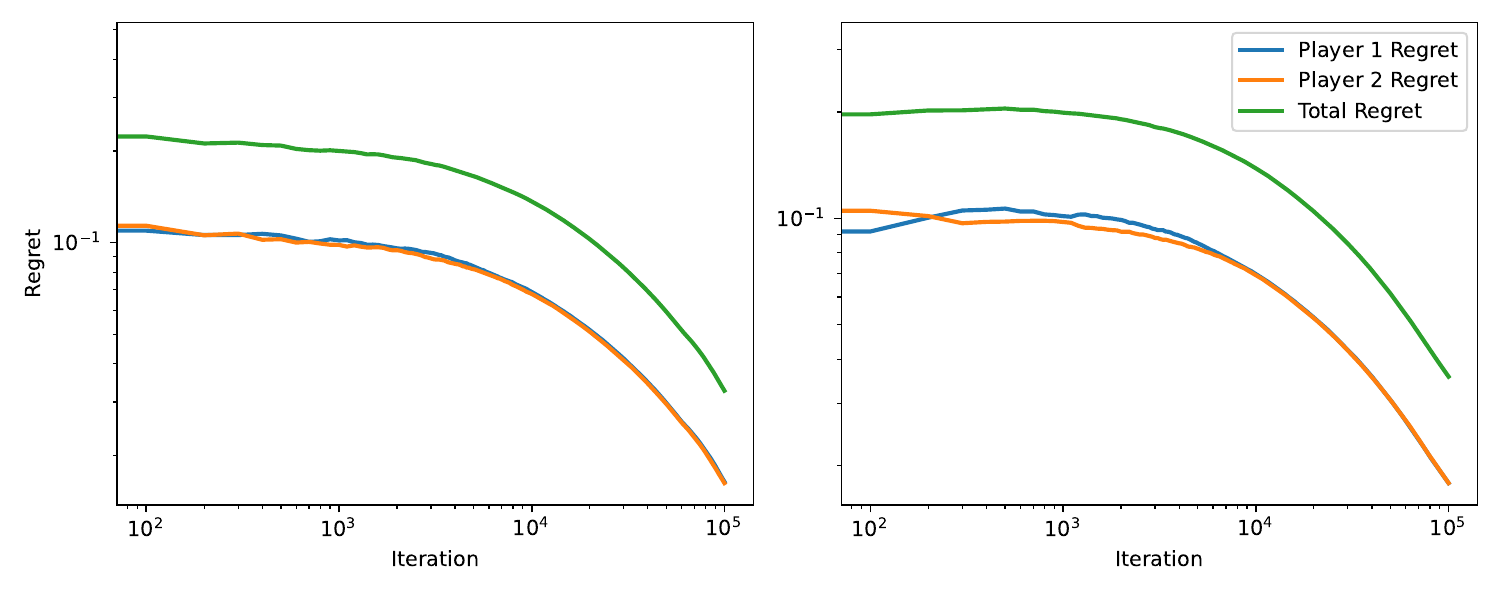}
    \caption{Total regret for the MWU allocations from Figure \ref{fig:adv_test} using a log-log scale. The horizontal axis in each panel represents the round of repeated play as detailed in Algorithm \ref{alg:meta}.}
    \label{fig:adv_regrets}
\end{figure}

All experiments are run on a machine with a dual-core (CPU) processor with $12$ GB RAM. In the experiments conducted to generate Figures \ref{fig:allocations_not_optimistic_fixed}-\ref{fig:adv_regrets} and Figure \ref{fig:uniform_regrets}, we use $\beta = 0.995$ and stop the algorithm after $T = 10^5$ iterations. In the experiment used for Figure \ref{fig:reg_vs_eq}, we use $\beta = 0.995$ and stop the algorithm after $T = 10^4$ iterations due to the increased time required to calculate the exact distance to equilibrium. In Table \ref{tab:timing}, we use $\beta = 0.95$ and stop the algorithm when the total regret is less than $0.05$. Total regret is calculated every $100$ rounds in every experiment. The code is available at \url{https://github.com/thomasvalles/Colonel_Blotto_MWU}. 

\section{Regrets for figures in main text}

\newpage

\section{MWU is a practical algorithm for solving Blotto}

We found that MWU offers an efficient way to find approximate Nash Equilibria in the Colonel Blotto game. Using the $0/1$ winning rule and $\beta = 0.95$, we are able to obtain a total regret of less than $0.05$ in just over $11$ seconds with $n_1= n_2 = 30, k = 20$ (see Table \ref{tab:timing} for exact times for different choices of $n_1, n_2$, and $k$). Numerical inaccuracies arising from the division of $\beta$ raised a number which scales with the cumulative loss of a strategy in Algorithm \ref{alg:partition} prevented us from achieving a total regret much less than $0.05$. We were able to consistently achieve a regret of around $0.01-0.04$ for all winning rules by increasing $\beta$ to $0.995$ and having the algorithm run for $10^5$ iterations. In general, this stopping condition and choice for $\beta$ was slower than when $\beta = 0.95$ and the algorithm ran until the total regret was less than $0.05$. Implementing different winning rules is relatively simple, though the choice of winning rule can drastically impact the time taken for the algorithm to terminate.

As an example, in Figure \ref{fig:allocations_not_optimistic_fixed}, we present average allocations output from the MWU algorithm with $k = 5$ battles and $n_1 = n_2 = 20$ soldiers for two sets of battle values. Keeping the winning rule fixed, the MWU allocations were similar across most initialization schemes. The only large deviations occurred in the first set of battle values (Figures \ref{fig:allocations_not_optimistic_fixed}-\ref{fig:regrets_not_optimistic_fixed}) when the uniform allocation initialization and proportional allocation initialization were used for the $0/1$ winning rule with $I_0 = 5000$. In these scenarios, the most valuable battle received around two more resources on average than the other initialization schemes and the regret was between two to three times as much as the regret of the other initialization schemes for the $0/1$ winning rule. We verify the convergence of our optimization procedure in Figure~\ref{fig:regrets_not_optimistic_rand}.

\subsection{Optimization details}

In line $4$ of Algorithm \ref{alg:blotto}, the historical loss matrix may be updated using one of two update rules:
\begin{itemize}
    \item Standard update: $L_i^{(t)}(j, m) = L_i^{(t-1)}(i, m) + \ell_j(m; p_{j}^{(\tilde{i}, t)})$
    \item Optimistic update:  $L_i^{(t)}(j, m) = L_i^{(t-1)}(i, m) + 2\cdot \ell_j(m; p_{j}^{(\tilde{i}, t)}) - \ell_j(m; p_{j, t}^{(\tilde{i}, t-1)})$
\end{itemize}
where $p_{j, t}^{(\tilde{i})}$ represents number of soldiers allocated to battle $j$ by player $i$'s opponent in round $t$. We found the performance of the standard update to be similar to that of the optimistic update (see Section \ref{sec:standard vs. optimistic}). All of the following experiments use the standard update.  
\begin{algorithm}
\caption{Compute Partition Function}
\label{alg:partition}
\hspace*{\algorithmicindent} \textbf{Input:} {Player $i$} \\
\hspace*{\algorithmicindent} \textbf{Output:} {Partition function $f_k(j)$}
\begin{algorithmic}[1]
\STATE{$f_1(j) \gets \beta^{L_i(1, j)}, \quad j = 0\dots n$}
\FOR{$k' = 2,\dots, k$}
\FOR{$n' = 0,\dots, n_i$}
\STATE{$f_{k'}(n') \gets \sum_{i = 0}^{n'} \beta^{L_i(k',j)}\cdot f_{k'-1}(n'-j)$}
\ENDFOR
\ENDFOR
\RETURN{$f$}
\end{algorithmic}
\end{algorithm}

\begin{algorithm}
\caption{MWU}
\label{alg:BlottoMWU}
\hspace*{\algorithmicindent} \textbf{Input:} {Player $i$} \\
\hspace*{\algorithmicindent} \textbf{Output:} {Strategy $s$ where $s(k')$ represents the number of soldiers allocated to battle $k'$}
\begin{algorithmic}[1]
\STATE{Get partition function $f$ from Algorithm \ref{alg:partition}}
\STATE{$r \gets n_i$}
\FOR{$k' = k, \dots, 2$}
\FOR{$y = 0,\dots, r$}
\STATE{$w_{k'}(y) \gets \beta^{L_i(k',y)}\cdot f_{k'-1}\left(r - y\right) / f_{k'}(r)$}
\ENDFOR
\STATE{$s(k') \gets Y \sim w_{k'}$}
\STATE{$r\gets r - b(k')$}
\ENDFOR
\STATE{$s(1) \gets r$}
\RETURN{$s$}
\end{algorithmic}
\end{algorithm}

\begin{algorithm}
\caption{Colonel Blotto MWU}
\label{alg:blotto}
\hspace*{\algorithmicindent} \textbf{Input:} {Regret tolerance $\epsilon$} 
\begin{algorithmic}[1]
\STATE{Initialize $L_1$ and $L_2$}
\WHILE{$\text{Regret} > \epsilon$} 
\FORALL{Player, $i$}
\STATE{Update $L_i$ }
\STATE{Obtain allocation from Algorithm~\ref{alg:mwu}}
\ENDFOR
\ENDWHILE
\end{algorithmic}
\end{algorithm}

We prove that our stopping criterion for the optimization is correct.
\begin{proof}[Proof of Proposition~\ref{prop:regret_opt}]
Denote the payoff matrices for players $1$ and $2$ as $X$ and $Y$. A strategy played by player $1$ in round $t$ is denoted by $x_t \in \R^{|S|}$, where $S$ is the set of feasible allocations ($|S| = {N + k - 1\choose N}$). $x_t$ is $1$ in one component, corresponding to allocation played by player $1$ in round $t$. $\bar{x} = \frac{1}{T} \sum_{t\in [T]} x_t$. Let $r_1$ and $r_2$ be the regrets of player $1$ and $2$ after $T$ iterations of the algorithm. $e_j$ is the unit vector with a $1$ in component $j$. The regret may formulated as:
\begin{equation}
    \frac{1}{T}\sum_{t = 1}^T x_t^T X \bar{y} = \max_j\, e_j^T X \bar{y} - r_1
\end{equation}
\label{eq:reg_a}
\begin{equation}
    \frac{1}{T}\sum_{t = 1}^T \bar{x}^T Y y_t = \max_j\, \bar{x}^T Y  e_j - r_2
\end{equation}
\label{eq:reg_b}

Replacing $Y$ with $-X$ in Equation \ref{eq:reg_b} gives us:
\begin{equation*}
    -\frac{1}{T}\sum_{t = 1}^T \bar{x}^T X y_t = -\min_j\, \bar{x}^T X  e_j - r_2
\end{equation*}
Summing this with Equation \ref{eq:reg_a} yields:
\begin{equation*}
    \min_s\, \bar{x}^T X e_s = \max_j e_j^T X \bar{y} - (r_1 + r_2)
\end{equation*}
Note that 
\begin{align*}
    \bar{x}^T X \bar{y} &\geq \min_s\, \bar{x}^T X e_s\\
    \implies \bar{x}^T X \bar{y} &\geq \max_j e_j^T X \bar{y} - (r_1 + r_2)
\end{align*}
We may similarly show that 
\begin{equation*}
    \bar{x}^T Y \bar{y} \geq \max_j \bar{x}^T Y e_j - (r_1 + r_2)
\end{equation*}
Hence 
\begin{equation*}
    \min\left(\max_j e_j^T X \bar{y} - \bar{x}^T X \bar{y}, \max_j \bar{x}^T Y e_j - \bar{x}^T X \bar{y}\right) \leq r_1 + r_2 
\end{equation*}
\end{proof}

\begin{figure}[!htbp]
    \centering
    \includegraphics[width = 10cm]{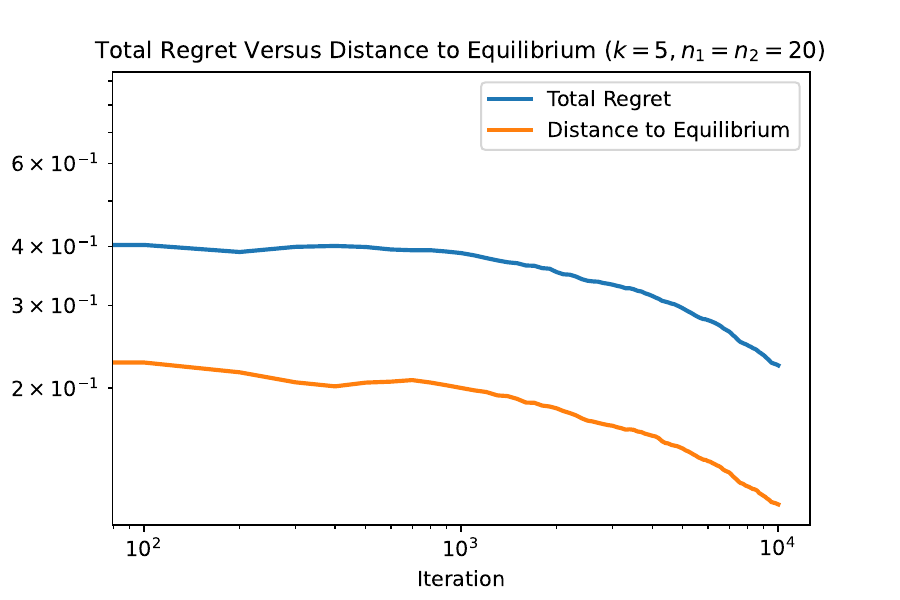}
    \caption{Total regret vs. distance to equilibrium up to $10^4$ iterations for the $0/1$ winning rule. Battle values are $v = [1, 2, 3, 5, 9]$. The horizontal axis in each panel represents the round of repeated play as detailed in Algorithm \ref{alg:meta}.}
    \label{fig:reg_vs_eq}
\end{figure}

This turns out to be a very loose bound, as demonstrated in Figure \ref{fig:reg_vs_eq}.

\section{Algorithm Speed}
\label{sec:algo_speed}
We demonstrate the algorithm's ability to find approximate equilibria solutions to the Colonel Blotto problem quickly. We can get within $5\%$ of the exact equilibrium solution for $k = 10$ battles and $n_1 = n_2 = 2500$ soldiers in $6$ hours and $16$ minutes.

\begin{table}[!htbp]
\centering
\begin{tabular}{|c|c|c|c|}
\hline
Battles & Soldiers & Time Standard Update (s) & Time Optimistic Update(s)  \\ \hline
10 &{[}20, 20{]} & 3.010 & 3.633 \\ \hline
10 &{[}20, 25{]} & 3.365 & 3.358 \\ \hline
10 &{[}20, 30{]} & 4.653 & 4.768 \\ \hline
10 &{[}25, 25{]} & 4.230 & 4.365 \\ \hline
10 &{[}25, 30{]} & 4.313 & 4.597 \\ \hline
10 &{[}30, 30{]} & 4.764 & 4.785 \\ \hline
15 &{[}20, 20{]} & 5.241 & 5.367 \\ \hline
15 &{[}20, 25{]} & 6.736 & 7.151 \\ \hline
15 &{[}20, 30{]} & 7.583 & 7.972 \\ \hline
15 &{[}25, 25{]} & 6.866 & 7.417 \\ \hline
15 &{[}25, 30{]} & 8.985 & 9.759 \\ \hline
15 &{[}30, 30{]} & 9.681 & 10.511 \\ \hline
20 &{[}20, 20{]} & 9.373 & 9.835  \\ \hline
20 &{[}20, 25{]} & 9.995 & 10.174 \\ \hline
20 &{[}20, 30{]} & 14.817 & 16.126 \\ \hline
20 &{[}25, 25{]} & 12.620 & 13.675 \\ \hline
20 &{[}25, 30{]} & 13.855 & 14.731 \\ \hline
20 &{[}30, 30{]} & 17.606 & 18.370 \\ \hline

\end{tabular}
\caption{Time required to achieve a regret of $5\%$ using the standard $0/1$ winning, no initialization, and randomized battle values. $\beta = 0.95$. Regret is calculated every $100$ rounds.}
\label{tab:timing}
\end{table}

\newpage 

\section{Standard Update vs. Optimistic Update}\label{sec:standard vs. optimistic}
\begin{figure}[!htbp]
    \centering
    \includegraphics[width = 14cm]{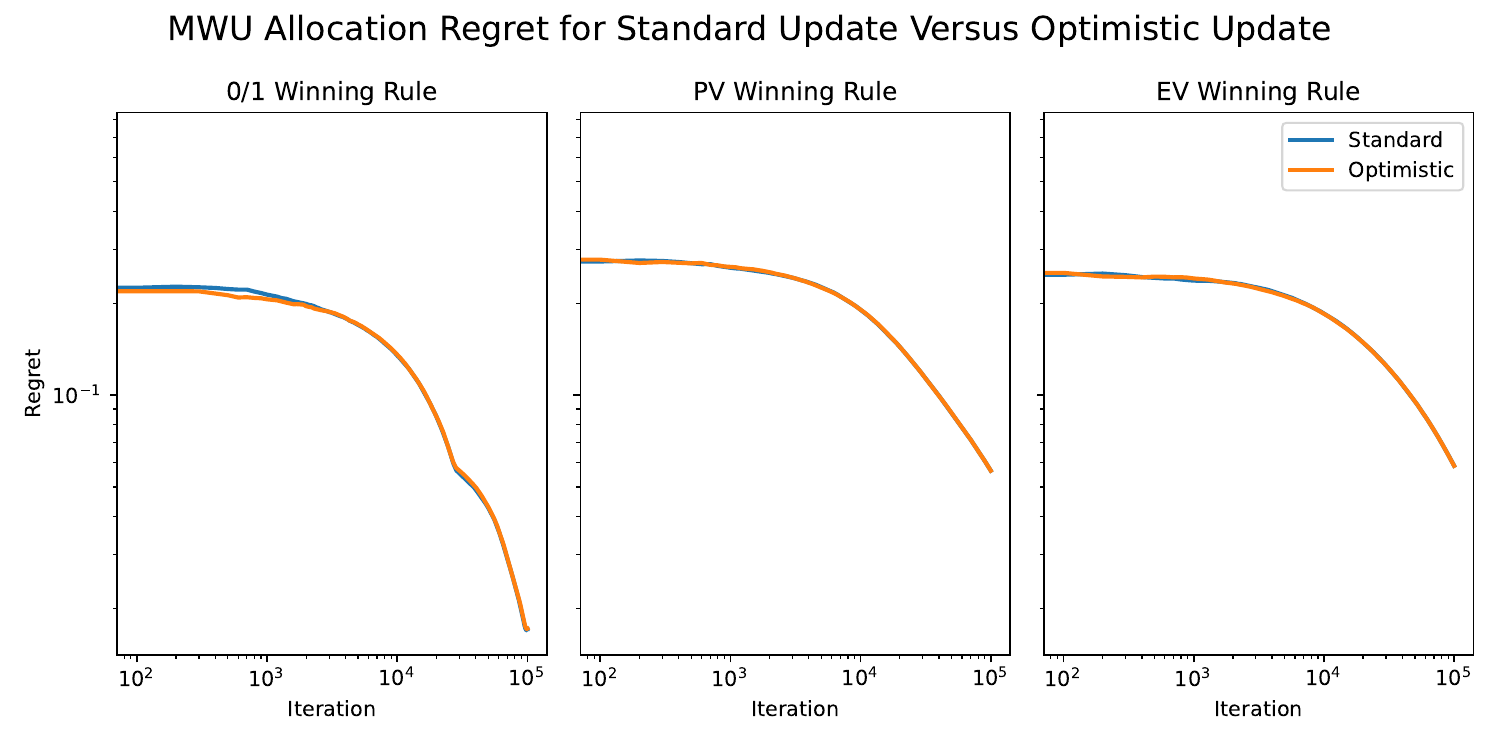}
    \caption{Total regret for optimistic and standard historical loss updates using the $0/1$ winning rule with $I_0 = 0$ for $k = 10$ uniformly valued battles and $n_1=n_2 = 20$ soldiers. $\beta = 0.995$. The horizontal axis in each panel represents the round of repeated play as detailed in Algorithm \ref{alg:meta}.}
    \label{fig:uniform_regrets}
\end{figure}

We address Question 1.13 in \cite{beaglehole2022sampling} to the negative - optimistic updates do not appear to give polylog regret when sampled rather than played determinstically. In Table \ref{tab:timing}, we see that using the optimistic update is slightly slower than the regular loss update rule. In Figure \ref{fig:uniform_regrets}, we plot the total regret over $10^5$ iterations of the algorithm using both the optimistic hedge and the standard historical loss update rule. The regret curves using the standard update and optimistic update are nearly indistinguishable. In all the following figures, the standard historical loss update is used.

\end{document}